\documentclass[twocolumn, switch]{article} 

\usepackage{preprint}

\usepackage[utf8]{inputenc}	
\usepackage[T1]{fontenc}	
\usepackage{xcolor}		

\usepackage{hyperref}
\hypersetup{colorlinks = true,
            linkcolor = purple,
            urlcolor  = blue,
            citecolor = cyan,
            anchorcolor = black} 

\usepackage{doi}
\usepackage{booktabs} 		
\usepackage{nicefrac}		
\usepackage{microtype}		
\usepackage{lineno}		
\usepackage{float}			

\usepackage{lipsum}		

\usepackage{newfloat}
\DeclareFloatingEnvironment[name={Supplementary Figure}]{suppfigure}
\usepackage{sidecap}
\sidecaptionvpos{figure}{c}

\usepackage{titlesec}
\titlespacing\section{0pt}{12pt plus 3pt minus 3pt}{1pt plus 1pt minus 1pt}
\titlespacing\subsection{0pt}{10pt plus 3pt minus 3pt}{1pt plus 1pt minus 1pt}
\titlespacing\subsubsection{0pt}{8pt plus 3pt minus 3pt}{1pt plus 1pt minus 1pt}

\newcommand\graycell{\cellcolor{lightgray}}

\usepackage{comment}
\usepackage{colortbl} 
\usepackage{scalefnt}
\usepackage{multirow}
\usepackage{subcaption,mwe} 
\usepackage{ amssymb } 

\title{Towards the Systematic Testing of  Virtual Reality Programs (extended version)}

\usepackage{authblk}

\author[1]{Stevão A. Andrade}
\author[2]{Fatima L. S. Nunes}
\author[1]{Marcio E. Delamaro}
\affil[  ]{Universidade de São Paulo}
\affil[1]{Instituto de Ciências Matemáticas e de Computação}
\affil[2]{Escola de Artes, Ciências e Humanidades}
\affil[  ]{\tt{\{stevao, delamaro\}@icmc.usp.br, fatima.nunes@usp.br}}

\begin{document}

\twocolumn[ 
  \begin{@twocolumnfalse} 


\maketitle

\begin{abstract}
\textbf{Abstract}

Software testing is a critical activity to ensure that software complies with its specification. However, current software testing activities tend not to be completely effective when applied in specific software domains in Virtual Reality (VR) that has several new types of features such as images, sounds, videos, and differentiated interaction, which can become sources of new kinds of faults. This paper presents an overview of the main VR characteristics that can have an impact on verification, validation, and testing (VV\&T). Furthermore, it analyzes some of the most successful VR open-source projects to draw a picture concerning the danger of the lack of software testing activities. We compared the current state of software testing practice in open-source VR projects and evaluate how the lack of testing can be damaging to the development of a product. We assessed the incidence of code smells and verified how such projects behave concerning the tendency to present faults. We also perform the same analyses on projects that are not VR related to have a better understanding of these results. The results showed that the practice of software testing is not yet widespread in the development of VR applications. It was also found that there is a high incidence of code smells in VR projects. Analyzing Non-VR projects we noticed that classes that have test cases tend to produce fewer smells compared to classes that were not tested. Regarding fault-proneness analysis, we used an unsupervised approach to VR and Non-VR projects. Results showed that about 12.2\% of the classes analyzed in VR projects are fault-prone, while Non-VR projects presented a lower fault-proneness rate (8.9\%). Regarding the application of software testing techniques on VR projects, it was observed that only a small number of projects are concerned about developing test cases for VR projects, perhaps because we still do not have the necessary tools to help in this direction. Concerning smells, we concluded that there is a high incidence in VR projects, especially regarding implementing smells and this high incidence can have a significant influence on faults. Finally, the study related to fault proneness pointed out that the lack of software testing activity is a significant risk to the success of the projects. 

\end{abstract}

\vspace{0.35cm}

  \end{@twocolumnfalse} 
] 



\section{Introduction}\label{sec:introducao}

Researchers have studied software faults extensively. These studies lead to the characterization of defects, both in the theoretical and practical contexts. Such characterization is essential to evaluate software testing techniques concerning their ability to reveal certain types of faults. These characterizations and taxonomies can provide guides for defining test approaches, which can support fault detection in specific software domains.

Technological advancement has led to the development of systems with new features such as images, sounds, videos, and differentiated interaction. Thus, technologies such as Virtual Reality (VR) have led to possibilities of creating three-dimensional environments with real-time interaction.

There are various techniques to obtain a greater sensation of immersion, according to the computational resources, equipment and systems used, vision, touch, and hearing experiences can be reproduced. Thus, in addition to simulating real situations, VR also allows users to depict and interact with hypothetical situations, involving static or moving virtual objects.

Despite the great benefits of adopting VR for the development of applications in various areas, it poses new challenges for verification, validation, and testing (VV\&T) activities. For example, VR software presents original software structures, such as scene graphs, which may represent new sources of defects for programs. These new challenges motivated the development of some approaches that aim to contribute to the quality assurance process of software in the context of VR.

As mentioned by Corrêa et al. \cite{Santos:2013}, there is interest in the literature on the subject. However, there is still no concept regarding systematized practices for conducting this activity. Studies have shown that the major problem remains in the difficulty to deal with test oracles, which is considered to be an open-ended research problem.

In general, the test activities for the VR domain are manually performed and mostly conducted only after the end of the development phase \cite{Correa:2018}. Such events generally support the generation of test requirements (functional and non-functional) which must be guaranteed before the product is delivered. The lack of studies that evaluate the cost of developing new techniques or using existing ones assess their effectiveness or even propose tools that can support their application, thus contributing to impact VV\&T activities in general and aggravate this scenario.

Regardless of the programming technology used, a primary development goal is to produce high-quality software. Consequently, VR also needs to be tested and vetted for quality. According to Neves et al. \cite{Neves:2018} there are series of conceptual challenges and key questions related to quality that need to be addressed when planning to apply software testing practices in new domains: \textit{What should be tested?, What does ``adequate testing" mean?, What is a failure in VR software?, Can we reuse something?, What is done nowadays?}

Systematic testing of the VR system must be based on fault models that reflect the structural and behavioral characteristics of VR software. Criteria and strategies for testing VR should be developed based on that fault model.

In this paper, we discuss whether new challenges for VV\&T of VR exist that require novel techniques and methods, or instead, we need new ways of combining and using existing approaches. We also try to evaluate how much the lack of VV\&T activities can negatively impact VR software development. To do so, we analyze the most popular open-source projects and categorize fault-proneness codes that could be mitigated by adopting VV\&T activities.

This paper is organized as follows: Section \ref{sec:related} present the related work and discusses how they differ from our work; Section \ref{sec:challenges} discusses the critical questions described above, as well as what testing approaches proposed in other domains could be reused for the VR domain; Section \ref{sec:experiment} presents an exploratory study to assess how much the lack of VV\&T activities can be prejudicial to open-source projects; Section \ref{sec:discussion} discusses the results of the experiment presented; Section \ref{sec:limitations} points out some limitations related to this study; finally the conclusions and future work are shown in Section \ref{sec:conclusion}.

\section{Related Work}\label{sec:related}

Although virtual reality studies date back a long time \cite{Boman:1995}, only recently have few studies addressed the development of VR applications from a software engineering perspective. The increase in community interest has emerged with the recent popularization of tools that have facilitated access, and consequently developers' interest in this technology, which is still considered as emerging.

Rodriguez and Wang \cite{Rodriguez:2017} present a survey about projects developed for the \textit{Unity} platform, highlighting the growth of the number of projects in recent years. Another highlight is the fact that despite the higher number of applications focused on games and entertainment, there has been an increase in the number of applications for other purposes, such as training and simulations.

Unlike our work, the paper does not analyze the content of the cataloged material in detail, and is limited to studying the growing trends, development involvement, favorite topics, and frequent file changes in these projects.

Ghrairi et al. \cite{Ghrairi:2018} conducted a study on the exploratory analysis of \textit{Github} projects and questions extracted from \textit{StackOverflow}, analyzing it from a software engineering point of view. The study demonstrates the current state of practice regarding the development of open-source VR applications, highlighting mainly the most used platforms and technologies. Moreover, the paper also discusses topics of interest for VR developers by analyzing the VR questions extracted from \textit{StackOverflow}.

The main results show the greater popularity of the \textit{Unity} and \textit{Unreal Engine} platforms as being the most popular among VR application developers. However, they also point out that more work needs to be done to better understand the VR requirements under a software engineering context, which is one of the points that our work seeks to elucidate.

From the perspective of software testing applications in the context of VR, we highlight the work produced by Corrêa et al. \cite{Correa:2018}, which presents a proposal for application software testing for VR applications. This study generates test data using specified requirements through a semi-formal language for VR application development. This approach moves in the direction pointed out by our study, which is the proposition of mechanisms that allow the systematization of the test activity for VR applications.

Karre et al. \cite{Karre:2019} conducted a year-long multi-level exploratory study to understand the various software development practices within VR product development teams in order to understand whether the traditional software engineering practices are still exercised during VR product development. The main results show that VR practitioners adopted hybrid software engineering approaches in VR product development and, in general, interviewed developers complain about the lack of software testing tools. An alternative presented to solve this problem is the involvement of consumers during the testing (pre-release versions) phase to understand the customer experiences and reactions.

Posteriorly Karre et al. \cite{Karre:2020} also assessed aspects related to tasks such as scene design, acoustic design, vergence manipulation, image depth, etc. which are specific to VR apps and hence require evaluation processes that may be different from the traditional ones. They presented a categorization that can support decision making in choosing a usability evaluation method for the future development of VR products. By using the categorization, the teams can use unique methods to improve the usability of their products, once depending on the industry (automobile, education, healthcare, etc.) the usability and the metrics of evaluation methods can change.

\section{Challenges and Issues}\label{sec:challenges}

The central concept in the VR literature is to define a more advanced form of user-computer interaction in real-time using synthetic three-dimensional multisensory devices \cite{Burdea94, LaValle:2019}. Objects can be presented, providing a sense that they are not the same as the user's perception when observing them in a limited area, such as a monitor screen.

VR can be classified according to the user's presence, as immersive and non-immersive. It is said to be immersive when the user has the sensation of being present within the virtual world using sensory devices (stereoscopic glasses or helmets with motion trackers) that capture their movements and behavior. Moreover, it is considered non-immersive when the user is partially transported to the virtual world via a monitor or projector \cite{Tori06}.

\begin{figure*}
    \centering
    \includegraphics[scale = 0.7]{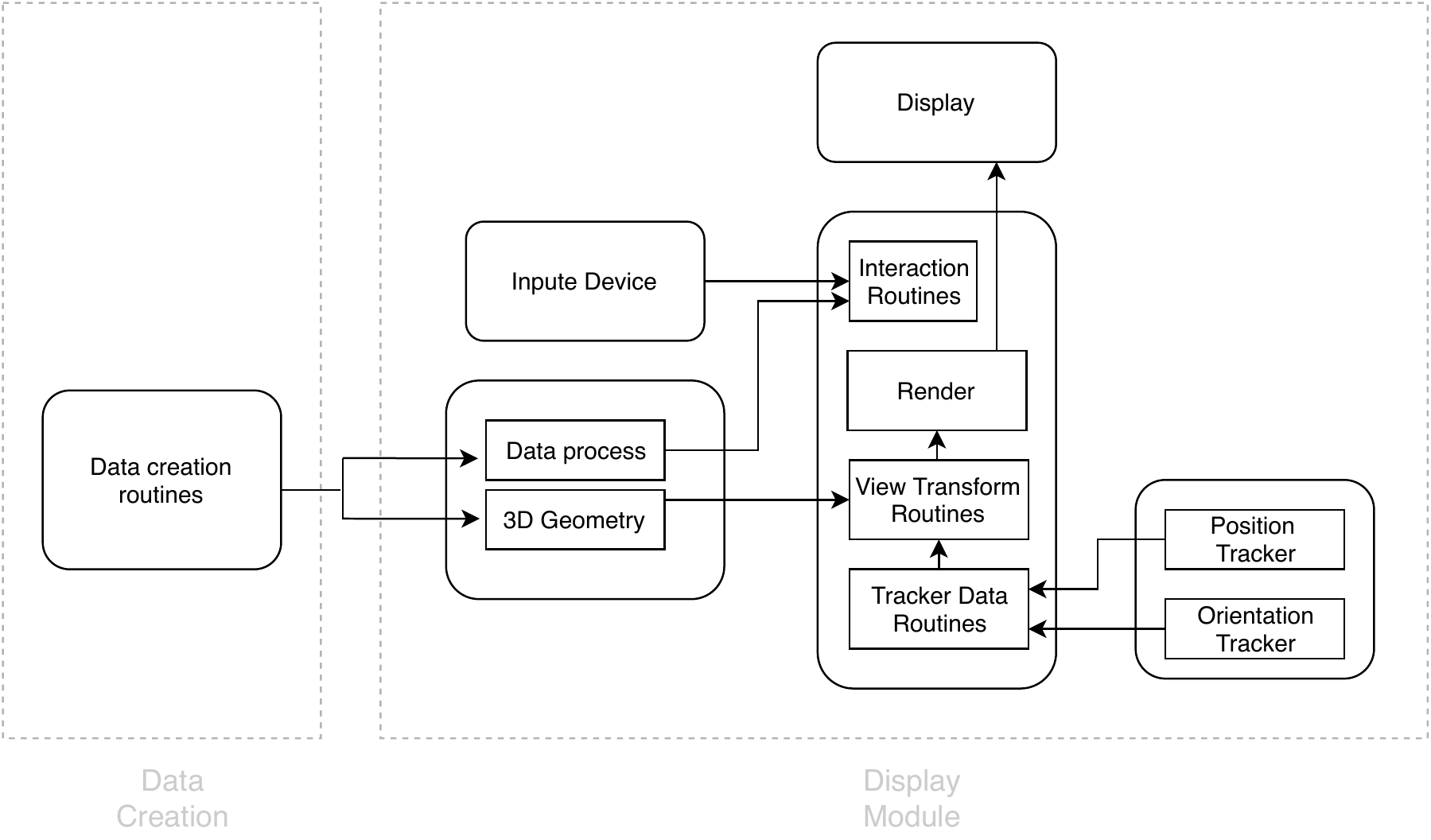}
    \caption{General VR application architecture (Source: own construction,  based on \protect\cite{Capilla:2004})}
    \label{fig:vr_arch}
\end{figure*}

According to Machado et al. \cite{machado_2003} and LaValle \cite{LaValle:2019}, immersion, and interactivity correspond to how the user interacts with the virtual environment, using devices that allow the manipulation of virtual objects. Several graphical libraries have emerged to help develop applications that reproduce requirements in a virtual environment; these libraries act as an abstraction layer providing information about the device's position or orientation, without the application knowing in which device the information is processed.

Despite the benefits of adopting VR for the development of applications in several areas, this poses new challenges for software quality assurance activities. For example, software developed for the context of VR has unique software structures, which may represent new sources of faults for the programs developed \cite{Takala:2017}. These new challenges have motivated the development of some approaches that aim to contribute to the quality assurance process of software in the context of VR.

Automating software testing activities is often a complicated and challenging process. The main tasks of this activity include organizing, executing, registering the execution of the test cases, and verifying the result of their execution.

In order to address these tasks, in the context of VR, some key points discussed in the next subsections should be understood.

\subsection{What should be tested?}\label{subsec:what_test}

Virtual reality systems use individual hardware devices to allow the interaction with the user and the system. The work of graphics engines is not the primary concern for VR application developers. Defining scene graphs for organizing 3D objects in a VR world, managing virtual users, controlling sensors for detecting events such as object collision and processing events for reacting to user inputs are some of the typical elements of VR systems that the developers should be concerned about \cite{Zhao:2009}.

According to Runeson \cite{Runeson:2006}, unit testing aims to test the smallest units that make up the system, thus ensuring the proper functioning of elements and easy-to-find programming faults, incorrect or poorly implemented algorithms, incorrect data structures, limiting the internal logic within the limits of the unit.

Figure \ref{fig:vr_arch} presents a general architecture of VR applications, as we can observe, VR applications differ from general programs by handling specific devices, and data structures used to represent the objects in a three-dimensional scene. Beyond represent various aspects of the real or imaginary world, such as geometric descriptions, appearance, transformations, behaviors, cameras, and lighting. Each of the properties above is created, aiming to represent objects present in a virtual environment, thus emerging new challenges related to how to test them.

By observing the organization of 3D object elements and assets in scene graphs, it seems that it needs a higher-level type of test. In general, because they are independent, they do not have an architecture correlation of the source code. Therefore, integration testing tends to be a more appropriate approach to be used. In integration testing, the main aim is to verify the communication between the units that make up the system.

\subsection{What does ``adequate testing'' mean?}

The solution to define this question: \textit{``What does adequate testing mean?"} is to apply test criteria, which consists of a set of rules for dividing and evaluating the valid input domain for the program being tested. A test criterion defines elements of a program that must be exercised during its execution, thereby guiding the tester in the process of designing the test cases for the system. A test requirement may be, for example, a software-specific execution path, a functionality obtained through specification, a mutation-based approach, etc \cite{Bertolino:2003}.

\begin{figure}[ht]
    \centering
    \includegraphics[scale = .43]{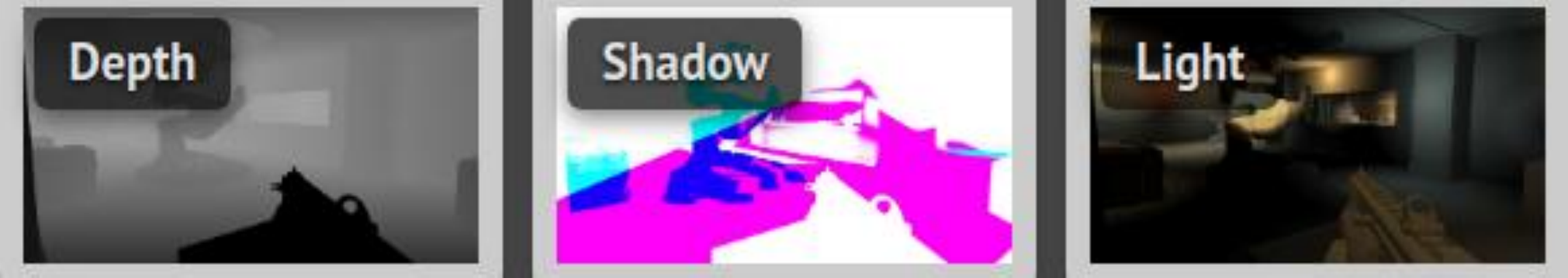}
    \caption{Example of scene rendering process (Source: own construction, based on \protect\cite{Adrian:2016})}
    \label{fig:scene_render_ex}
\end{figure}

As pointed out in the previous question, due to the different structures existing in VR applications, it is difficult to define which aspects of a VR application should be considered when designing a test routine for VR applications. For example, the Figure \ref{fig:scene_render_ex}  shows that for a single frame existing in a 3D scene, there are different layers that have different aspects that can be taken into account when testing the application.

Corrêa et al. \cite{Santos:2013} presented a set of studies that deal with the application of software testing techniques to programs in the VR context, showing that there is an interest in the literature on the subject, however, there is still no concept regarding systematized practices for the activity.

Due to the lack of defined requirements, it is not easy to identify test adequacy criteria for VR systems. How can we decide that testing is enough? This question needs to be adapted to the context.

\subsection{What is a failure in a VR software?}

In the context of VR applications, the testing activity hinges on the difficulty of systematizing how the behavior of a test case can be measured. This difficulty is described in the literature as ``test oracle problems" and it appears in cases where traditional means of measuring the execution of a test case are impractical or are of little use in judging the correction of outputs generated from the input domain data \cite{Rapps:1985, Barr:2015}.

Test oracles deal with a primary issue in software testing activities - deciding on program correctness based on predetermined test data. In general, the test oracle accesses a data set needed to evaluate the correctness of the test output. This data set is taken from the specification of the program under testing and should contain sufficient information to support the final decision of the Oracle \cite{Oliveira14}. 

It is possible to explore a wide range of faults in the context of VR software. It is possible to verify the strictness of specification based on scene graphs concepts, in addition to specific features related to the virtual environment, it is also possible to verify the behavior of objects and the actions performed by multiple devices. 

As can be seen, in addition to traditional source code routines, several characteristics need to be taken into account when testing and the definition of what can be considered a failure is an important step for each of the described aspects above can be analyzed correctly during the testing activity.

\subsection{Can we reuse something?}

General tools such as capturing and replaying can be used, but they offer a shallow level of abstraction. Thus, any small change to the system will result in the fact that the tests should be redone \cite{Leotta:2013}. Therefore, using capture and replay tools cannot be used when the system is in development.

From a unit test point of view, we can still reuse a traditional approach in which we can quickly gauge the expected output to a method execution, ensuring that the smallest units of the VR system have been sufficiently tested against their specifications.

Regarding integration testing which is expected to handle new kinds of elements (3D objects, assets, behaviors, etc.) the literature review shows that we still need better-systematized practices for this activity \cite{Santos:2013}.

\subsection{What is done nowadays?}

Almost all 3D applications require some common features. Therefore, developers tend to use platforms that provide these features out-of-the box. Using game engines is one of the most popular techniques among developers due to the fact it helps produce the systems, besides speeding up the development process.

Recently popular game engines, such as \textit{Unity3D}\footnote{\url{https://docs.unity3d.com/Manual/testing-editortestsrunner.html}} and \textit{Unreal Engine}\footnote{\url{https://docs.unrealengine.com/latest/INT/Programming/Automation}}, released their own set of testing tools, which allows developers to produce automated testing during the development phase of the system which can substantially increase the stability of the product developed.

Despite the existence of tools, it still does not provide observable test criteria, perhaps due to the lack of studies that propose, experiment and validate effectively applicable approaches, which can be additionally repeatable, documented and do not rely only on the tester’s creativity.

\section{Do We Really Need to Test Virtual Reality Software?}\label{sec:experiment}

Considering popularizing VR application development, we are interested in understanding, from the software engineering point of view, how the development process of these applications is currently conducted. We are especially interested in software testing practices in the development process of such applications in order to address what kinds of malfunctions the lack of test practice can lead to.

One of the most used approaches to quantify quality attributes in software projects is the evaluation of source code metrics. Source code metrics are a significant component for the software measurement process and are commonly used to measure fault proneness and improve the quality of the source code itself \cite{Palomba:2014}.

\begin{figure*}[t]
    \centering
    \includegraphics[scale = 0.6]{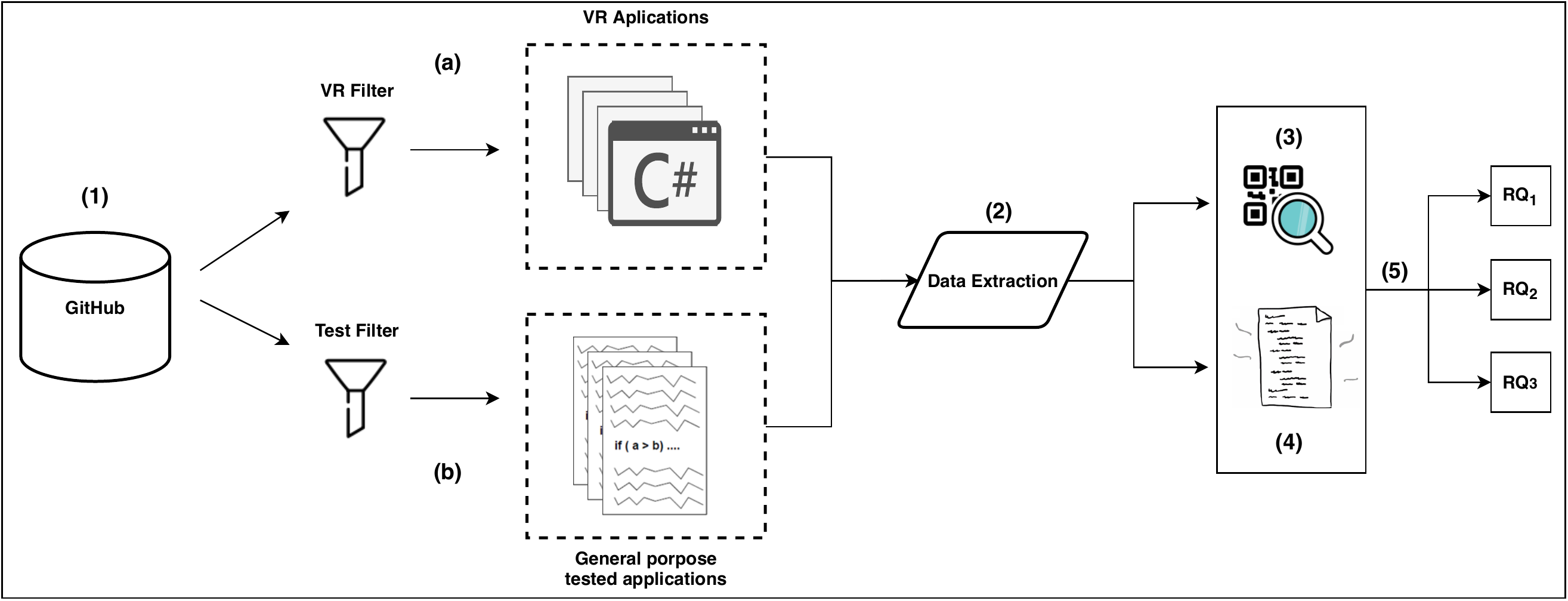}
    \caption{Steps taken to carry out the study (Source: own construction)}
    \label{fig:study_design}
\end{figure*}

Another factor that can be exploited to evaluate code quality is to identify anti-patterns since some studies show that there is a correlation with fault proneness \cite{Khomh:2012}. Therefore, these are two aspects that are taken into account in the evaluation carried out in our study to investigate the quality aspects of the code in the context of software testing.

Ghrairi et al. \cite{Ghrairi:2018} made an exploratory study on \textit{Github} and \textit{Stack Overflow} in order to investigate which are the most popular languages and engines used in VR projects. According to their results, the most popular language for VR development is C\#, and Unity is the most used game engine during VR application development. Thus, we focus our analyses targeting these characteristics.

\subsection{Overview of the study}

We formulated the following research questions regarding the quality analysis goal of VR projects.

\begin{itemize}
    
    \item $\mathbf{RQ_1}:$ \textit{``How does testing happen in open-source VR software systems?"} We focus on understanding how testing practices are being applied in open-source VR projects. 
    
    \item $\mathbf{RQ_2}:$ \textit{``What are the distribution of architecture, implementation and design smells in VR projects?"} We investigated the distribution of smells to find out whether there is a set of code smells that occur more frequently in VR systems.
    
    \item $\mathbf{RQ_3}:$ \textit{``Can we draw a relationship between code metrics and fault proneness?"} It is commonly believed that code metrics and fault-proneness, i.e., if a set of code metrics reaches a predefined threshold, it is very likely that the project could also have some faults. We investigate this using an unsupervised defect prediction approach. 
 
\end{itemize}

\begin{table*}[ht]
  \scalefont{0.85}
  \centering
  \caption{Characteristics of the repositories used in the experiment}
    \begin{tabular}{l|r|r|r|r|r|r}
    \hline
    \multicolumn{1}{c|}{\multirow{2}[4]{*}{Features / Characteristics}} & \multicolumn{2}{c|}{Small size ( 1 $\thicksim$ 80 classes)} & \multicolumn{2}{c|}{Medium size (80 $\thicksim$ 200 classes)} & \multicolumn{2}{c}{Large size ( 201+ classes)} \\
\cmidrule{2-7}          & \multicolumn{1}{c|}{\textbf{VR}} & \multicolumn{1}{c|}{\textbf{non-VR}} & \multicolumn{1}{c|}{\textbf{VR}} & \multicolumn{1}{c|}{\textbf{non-VR}} & \multicolumn{1}{c|}{\textbf{VR}} & \multicolumn{1}{c}{\textbf{non-VR}} \\
    \hline
    Nº  C\# Classes & 31.5 & 30.2  & 121.7  & 126.4  & 374.8  & 4320  \\
    \hline
    Nº Branches & 1.45  & 1.7   & 1.7   & 2.5   & 2.56  & 11.4 \\
    \hline
    Nº Commits & 78.4  & 45.5  & 48.8  & 245   & 66.2  & 8931 \\
    \hline
    Nº Contributors & 2.6   & 1.4   & 2.5   & 2.2   & 2.9   & 47.1 \\
    \hline
    Nº Forks & 34.7  & 1.9   & 10.9  & 1     & 76.6  & 82.4 \\
    \hline
    Nº Subscribers & 15.8  & 2     & 8.1   & 2.5   & 22.3  & 176.8 \\
    \hline
    Nº Stars & 111.6 & 4     & 23.8  & 0.7   & 187.8 & 89.4 \\
    \hline
    \end{tabular}%
  \label{tab:github_stats}%
\end{table*}%

\subsection{Study Design}

The process of selecting open-source projects consists of a systematized search, on \textit{Github}, using the keywords ``virtual reality" and ``VR". With the objective of drawing a more specific profile, the search focused only on projects developed for the \textit{Unity} platform, since it has emerged as the most popular VR development platform due to its extensive documentation \cite{Ghrairi:2018}.

Our primary aim is to explore virtual reality projects from both project and source code perspectives. To do so, we cataloged and analyzed a total of 151 open-source projects, available in \textit{Github}. Some of the projects could not be analyzed due to either missing external dependencies or custom-build mechanisms (i.e., missing standard C\# project files), thus we were able to analyze a total of 119 projects.

In order to draw a picture concerning research questions $RQ_2$ and $RQ_3$, we also cataloged a set of general (Non-VR) open-source projects, which have similar characteristics (same C\# programming language). The goal is to try to compare the information observed in the VR application with Non-VR application. Therefore, we catalog a total of 177 Non-VR projects. After an individual process analysis, we removed duplicated projects and projects that had missing external dependencies or custom-build mechanism. In the end we achieved a total of 107 Non-VR projects able to be used in our experiment.

Since our goal is to analyze which impacts the lack of software testing practice can cause on VR projects, concerning Non-VR projects, we will use only data related to classes that have been properly tested.

Thus, for all the experiments, two types of projects were gathered. The first is a set of VR projects. The second is a set of general-purpose projects that hold some unit test code used to evaluate their modules.

The purpose of the first step is to manually evaluate VR projects in order to understand how much open-source VR application developers are concerned regarding testing practices for this specific domain.

Based on the observation of the results of the initial analysis, the second step is to assess how much the lack of software testing activity can contribute to the construction of codes that become more difficult to maintain over time. For this, an analysis is made regarding the presence of code smells and the results obtained in VR projects are compared with the results obtained in general-purpose projects.

Finally, the last step aims to assess how much code metrics are capable to point about fault proneness. The goal is to evaluate the projects and observe how the VR projects and projects that have been tested behave, in order to understand whether test practice can contribute to a given code snippet being less fault-prone regarding untested code snippets.

The procedure carried out during the execution of the experiment is described in Figure \ref{fig:study_design} and consists of the following steps:

\begin{enumerate}

    \item Search the \textit{Github} platform for projects with the targeted characteristics.

    \begin{enumerate}
    
        \item VR projects that were developed using the Unity platform.

        \item General purpose projects that were developed using the C\# programming language and have software testing practices in their composition.

    \end{enumerate}
    
    \item Data extraction related to testing practice in VR application projects.

    \item Code metrics and code smells information extraction for all projects.

    \item Use of code metrics to calculate fault-proneness in all projects.

    \item Summary of the results and answer the research questions.
    
\end{enumerate}

\subsection{Overview of Projects}

One of the biggest concerns when carrying out experiments using open-source artifacts is related to the representativeness of the artifacts and whether therefore the results found can be properly generalized \cite{Hillenbrand:2005}.

With it in mind, we were careful to try to select the largest number of study objects, in such a way that they could meet a wide variety of characteristics and not be limited to toy examples. Table \ref{tab:github_stats} summarizes the information on the artifacts used in the experiments developed in this paper. The data related to the main characteristics of the repositories  are presented in the form of average, by classifying the repositories among small projects (with up to 80 classes), medium (with projects that have between 80 and 200 classes) and large projects with a number of higher classes to 200.

Below we present the main features cataloged in the repositories used:

\begin{itemize}

	\item Branches - correspond to branches used to implement new features or correct faults, in such a way that the unstable code is not merged into the main project files.

	\item Commits - commits refer to the act of submitting the latest source code changes to the repository and making these changes part of the main version of the repository.

	\item Contributors - refers to the number of people outside the main project development team who have contributed or wish to contribute to changes to the project.

	\item Forks - copies from a repository. Forking a repository allows third parties to perform experiments on the source code without affecting the original project.

	\item Subscribers - number of people who are registered to receive notifications related to updates on activities and changes made to the project.

	\item Stars - indicates the number of people who have defined markers in the project so that they can follow what is happening with the project in the future ( according to Kalliamvakou et al. \cite{Kalliamvakou:2014} this metric is commonly used to measure the popularity of repositories).
	
\end{itemize}

In order to provide a view of the general scope of the artifacts used, Table \ref{tab:projects_overview} presents information about the general characteristics of all projects.

\begin{table}[ht]
    \centering
    \caption{General characteristics of the analyzed projects}
    \begin{tabular}{l|r|r}

        \hline
        \multicolumn{1}{c|}{\textbf{Attributes}}   & \multicolumn{1}{c|}{\textbf{VR}} & \multicolumn{1}{c}{\textbf{Non-VR}} \\ 
        \hline

        Projects                    &   119          &  107      \\
        Number of tested classes    &   63           &  4,186      \\
        Number of classes (total)   &   21,508       &  21,563      \\
        Lines of code (C\# only)    &   2,314,522    &  2,455,766    \\

        \hline 

    \end{tabular}
    \label{tab:projects_overview}
\end{table}

Aiming to estimate non-functional requirements used to evaluate the performance of a system, such as software quality attributes of the projects, and to give an overall view of the projects analyzed, we computed, according to an object-oriented design \cite{Chidamber:1994}, a set of metrics that are summarized in Table \ref{tab:projects_metrics}.

\begin{table}[ht]
  \scalefont{0.7}
    \centering
    \caption{Metrics of the analyzed projects}
    \begin{tabular}{l|r|r|r|r}

        \hline
        \textbf{Metric}    &    \textbf{VR}    &    \textbf{Average}    &    \textbf{Non-VR}    &    \textbf{Average} \\ 
        \hline
        
        Number of Children            &    3,811      &    0.17    &    716
    &    0.17    \\
        Number of Fields              &    102,785    &    4.77    &    7,062
    &    1.68    \\
        Number of Methods             &    107,516    &    4.99    &    14,374
    &    3.43    \\
        Number of Properties          &    24,395     &    1.13    &    67
    &    0.01    \\
        Number of Public Fields       &    49,793     &    2.31    &    1,368
    &    0.32    \\
        Depth of Inheritance Tree     &    4,072      &    0.18    &    1,278
    &    0.30    \\
        Number of Public Methods      &    65,624     &    3.05    &    9,582
    &    2.28    \\
        Lack of Cohesion of Methods   &    34,197     &    1.58    &    7,958
    &    1.90    \\
        Weighted Method per Class     &    190,658    &    8.86    &    21,959
    &    5.24    \\ \hline

    \end{tabular}
    \label{tab:projects_metrics}
\end{table}

All the information about the projects, the data used for plotting the tables and graphs, as well for the discussion, are available in the experiment repository\footnote{\url{https://github.com/stevao-andrade/ACL_defect_prediction}}.

\section{Results and Discussion}\label{sec:discussion}

In this section, we address the research questions and further discuss the results obtained from the analysis and our observations with the empirical study.

\subsection{How is VR software tested?}

Existing software testing techniques seek to establish rules to define test case sets that exercise the testing requirements needed to eliminate software faults when they exist. Testing techniques and criteria are of paramount importance in selecting test cases because the tester can create more effective test cases, and thus reduce the number of test cases, ensuring that specific pieces of software are exercised by the tests. Testing techniques differ from one another in the source of information used to create the test requirements.

Regarding the first research question ($\mathbf{RQ_1}$) of the study, which aims to understand the question: \textit{``How does testing happen in open-source VR software systems?"}, the 119 projects were manually evaluated and it was found that only 6 VR projects  (\textit{Bowlmaster - 53 tests, CameraControls - 60 tests, GraduationGame - 15 tests, MiRepositorio\_VRPAD - 11 tests, space\_concept - 11 tests, UnityBenchmarkSamples - 4 tests}) are concerned with the software testing practices, including a total of 154 unit test cases,  to evaluate the projects' functionalities.

Despite the existence of unit testing, we were unable to calculate information regarding testing criteria, such as code coverage, since \textit{Unity} does not provide an out-of-the box solution to code coverage (till the present time) and \textit{Unity} uses its own fork of \textit{Mono} \cite{Mono}, which makes it impractical  to use other C\# coverage tools \cite{Haas:2014}.

Based on the information collected, it can be observed that from  the 119 analyzed projects, only 6 (5.04\%) have some software testing activity, and even the projects that have test cases, do not present many tests that can ensure that the main functionalities of the applications were adequately tested.

Another interesting aspect observed when analyzing the 6 projects that have test cases is the fact that the \textit{Bowlmaster} project is part of a popular course, with more than 334,000 students enrolled \footnote{\url{https://github.com/CompleteUnityDeveloper/08-Bowlmaster-Original}}, which aims to teach VR application development practices. This demonstrates that there is a concern on the part of educators regarding the awareness that testing activity is a determining factor in the software development process and it is expected that students are capable of apply concepts related to testing practice software in their future projects.

Concerning $RQ_1$ we came to the conclusion that there is not yet consensus regarding the application of software testing practices for VR applications and this motivated us to explore the next research questions. These results are in agreement with the most recent papers in the literature.  Karre et al. \cite{Karre:2019} conducted an empirical study of VR practitioners to learn the current practices and challenges faced in industry. The software testing related results points out to the absence of adequate tools, as well as uncertainty about how to test the VR app apart from conducting a standard field evaluation. As a consequence, this lack of usability evaluation methods and automated testing tools tend to cost a lot of time to release a VR product.

One possible explanation is due to the challenges of systematizing how the behavior of a test case can be measured in the context of VR programs. This difficulty is described in the literature as a ``test oracle problem" and arises in cases where traditional means of gauging the execution of a test case are impractical or are of little use in judging the correctness of outputs generated from input domain data \cite{Rapps:1985, Barr:2015}. Test oracles deal with a primary issue in software testing activities - deciding on program correctness based on predetermined test data. In general, the test oracle accesses a data set needed to evaluate the correctness of the test output. This data set is taken from the specification of the program under testing and should contain sufficient information to support the final decision of the Oracle \cite{Oliveira14}.

It should be emphasized that building a project with good test cases is not an easy task. Testing requires discipline, concentration, and extra effort \cite{Kasurinen:2010}. As a reward, the code presents a set of characteristics such as cleanliness, easy-to-maintain, loosely coupled, and reusable APIs. Besides the fact that testable code is easier to understand, maintain and extend.

In order to understand the risks and advantages of these characteristics and to accurately answer $RQ_2$ and $RQ_3$,  in the next sessions, we compare the difference between the VR projects and Non-VR projects concerning code smells  and fault-proneness distribution.

\subsection{Distribution of Code smell}

Observing the lack of software testing practice in all the other projects, we decided to investigate how this practice is reflected within the projects. To do so, we decided to measure the incidence of code smells \cite{Fowler:2018} within the projects investigated. This leads us to the second research question ($\mathbf{RQ_2}$) presented: \textit{``What are the distribution of architecture, implementation, and design smells in VR projects?"}.

In general, the presence of code smells in software projects indicates the presence of quality problems. For instance, one of the most well-known code smell, God Class, is defined as a class that knows or does too much in the software system. God Class is a strong indicator of possible problems because this component is aggregating functionality that should be distributed among several others components, therefore increasing the risk of software faults \cite{Hall:2014}. Such problems directly impact features such as maintainability and contribute to make it difficult for software to evolve.

To perform the evaluation discussed in this section we started from an assumption that projects that do not have test cases in their composition tend to have a lower code quality opposed to projects that have been tested, considering developers probably may not have observed aspects that would be capable of triggering unexpected behavior in the application, besides the fact the smaller the number of bugs in the system, the higher the quality related to a given project.

In order to better understand what is related to the lack of tests, we compared the results obtained in the VR projects with the results obtained in Non-VR applications, which have well-defined test cases within the projects.

To better understand $\mathbf{RQ_2}$,  we identified three different types of code smells in the projects:

\begin{itemize}
    
    \item \textbf{Architecture smells}: focus on identifying points of interest for possible structural problems that can negatively contribute and hamper activities such as debugging and refactoring, as well as increasing the cost for fault correction and refactoring, due to the characteristic of increasing the complexity of the software, when present \cite{Mo:2015}.

    \item \textbf{Implementation smells}: code smells or implementation smells were first introduced by Fowler \cite{Fowler:2018} and seek to establish a concept to classify shortcomings in object-oriented design principles. This class of smells covers principles such as data abstraction, encapsulation, modularity, hierarchy, etc.

    \item \textbf{Design smells}: are specific types of structures that may indicate a violation of a fundamental principle, which can impact aspects of design quality \cite{Suryanarayana:2014}.

\end{itemize}

In order to calculate the distribution of the code smells previously described within the projects, we use the \textit{Designite} tool \cite{Sharma:2016}. The smells were classified according to the number of occurrences in the analyzed classes and percentage distribution. The data is presented in Tables \ref{tab:architecture_smells}, \ref{tab:implementation_smells} and \ref{tab:design_smells}.

It is worth mentioning that test case classes were not taken into account for this smell classification, once our initial target was to measure the quality aspects of the source code classes. Besides that, smells in software test codes require a whole different classification approach \cite{Tufano:2016}.

\subsubsection{Architecture smells results}

It can be observed that in Table \ref{tab:architecture_smells}, among the VR projects, there is a low incidence of architecture smells, with only three types (ACD, AFC, and AGC) presenting a percentage of occurrence between 0.93 \% and 1.70 \%. Observing the Non-VR projects, it can be observed that this category of smells had a lower incidence compared to VR projects. The AUD, AGC and AFC smells showed the highest occurrence rates, with percentages between 0.26\% and 0.57\%.

VR project behavior can be justified due to the fact that within the \textit{Unity} platform, although an object-oriented language (C\#) is mostly used, the development model is considered a component-based programming approach. This approach focuses on the separation of concerns regarding the features to be developed in the system.

\begin{table}[ht]
  \scalefont{0.73}
  \centering
  \caption{Description of the detected architecture smells and their distribution}
    \begin{tabular}{l|l|r|r|r|r}
    \hline \textbf{ID} & \textbf{Smell} & \multicolumn{2}{c|}{\textbf{VR}} & \multicolumn{2}{c}{\textbf{Non-VR tested}} \\ \hline
    
    AAI   &   Ambiguous Interface        & 29    & 0.13\% & 1     & 0.02\% \\
    ACD   &   Cyclic Dependency          & 212   & 0.99\% & 6     & 0.14\% \\
    ADS   &   Dense Structure            & 3     & 0.01\% & 2     & 0.05\% \\
    AFC   &   Feature Concentration      & 366   & 1.70\% & 24    & 0.57\% \\
    AGC   &   God Component              & 201   & 0.93\% & 12    & 0.29\% \\
    ASF   &   Scattered Functionality    & 84    & 0.39\% & 8     & 0.19\% \\
    AUD   &   Unstable Dependency        & 78    & 0.36\% & 11    & 0.26\% \\ \hline
    
    Std Dev &  \multicolumn{1}{c|}{\graycell}  & 118.34  &  \multicolumn{1}{c|}{\graycell}  & 7.17 & \multicolumn{1}{c}{\graycell} \\ \hline
    
    Average &  \multicolumn{1}{c|}{\graycell}  & 139.0   & \multicolumn{1}{c|}{\graycell}             & 9.14   & \multicolumn{1}{c}{\graycell} \\ \hline

    \end{tabular}
  \label{tab:architecture_smells}
\end{table}

Among the main advantages of a component-based programming approach, we can point out the high reuse capacity of the developed components due to the low coupling characteristics of the components that make up the systems.

Despite Non-VR applications presenting lower rates of architecture smells, it mainly shows a higher incidence of smell AFC. This smell occurs when a component performs more than one architectural concern/feature. This can be explained due to the programming model adopted. A large part of Non-VR projects corresponds to web applications, which typically use a Model-View-Controller (MVC) standard for application development. As shown by Aniche et al. \cite{Aniche:2018}, systems that adopt such architecture can be affected by types of poor practices that lead to the apparition of such a smell.

From a software testing point of view, the lower rate of architecture smells can be considered as a decisive successful factor, since the low dependence between modules is a characteristic that facilitates the application of unit tests \cite{Aniche:2013}. In general, when it is necessary to communicate with other units of code, sometimes \textit{stubs} or \textit{mock} objects are used to represent this communication. A huge benefit of this approach is that by lower coupling the system it is possible to reproduce complex scenarios more easily.

Despite the advantages, it is important to keep in mind that there are some threats related to the use of a component-based approach in the context of integration testing, which as discussed in subsection \ref{subsec:what_test} must be one of the characteristics prioritized in the context of testing VR applications. The main problems are related to the use of components produced by third parties since in general they work as a black box and it is necessary to trust their correct functionality. An example of this model is the components asset store available in the Unity platform\footnote{\url{https://assetstore.unity.com/}}.

In order to better understand the presented results, regarding each of the classes of smells analyzed, we verified if there is, in fact, a statistical difference between the presence of smells between groups of classes that were not tested and groups of classes that were tested during its development process. Therefore, due to the low number of smell types for each category (architecture, design and implementation), and since we can not guarantee that the data collected departs from a normal distribution, we applied the Mann-Whitney test \cite{Fay:2010} to verify whether there is a statistical difference between the presence of smells for each category of smells evaluated.

The null hypothesis ($H_0$) of the Mann-Whitney test indicates that \textit{``The distribution of the variable in question is identical (in the population) in the two groups"}, that is, there is no difference in the presence of smells between classes that have not been tested and classes that have been tested and the alternative hypothesis ($H_1$) indicates that \textit{``The distributions in the two groups are not the same"}, therefore, there is a statistical difference between the incidence of smells for classes that were not tested against classes that were tested.

Considering the value of alpha = 0.05, which comprises the complement of the margin of a confidence level of 95\%, for the architecture smells, $H_0$ with a p-value = 0.00760 could be rejected. Thus, it indicates that there is a statistical difference between the presence of smells when comparing architecture smells in classes that were not tested against classes that were tested.

Using a descriptive analysis, obtained by analyzing the number of occurrences of each type of smells, it could be observed that classes that were not tested tend to present a higher rate of architecture smells in relation to classes that were tested.

\subsubsection{Implementation smells results}

It can be observed that, different from the architecture smells, in Table \ref{tab:implementation_smells}, we can identify a high rate of implementation smells in the VR projects. We highlight ILI, ILS, and IMN, which had a percentage of occurrence of 31.81\%, 55.55\%, and 117.46\%, respectively.

Although it does not pose a direct risk to the source code produced, smell ILI may be an indicator that something can be revised/refactored. A very long identifier may be an indication that there is a need for too much text to distinguish/identify variables and in some instances, this may indicate that the programmer may not be using the most suitable data structure to represent it.

\begin{table}[ht]
  \scalefont{0.69}
  \centering
  \caption{Description of the detected implementation smells and their distribution}
    \begin{tabular}{l|l|r|r|r|r}
    \hline \textbf{ID} & \textbf{Smell} & \multicolumn{2}{c|}{\textbf{VR}} & \multicolumn{2}{c}{\textbf{Non-VR}} \\ \hline

    ICM   &   Complex Method           &  1,812     &   8.42\%     &  9   &  0.22\% \\
    ICC   &   Complex Conditional      &  684       &   3.18\%     &  14  &  0.33\% \\
    IDC   &   Duplicate Code           &  9         &   0.04\%     &  1   &  0.02\% \\
    IECB  &   Empty Catch Block        &  150       &   0.70\%     &  5   &  0.12\% \\
    ILM   &   Long Method              &  583       &   2.71\%     &  9   &  0.22\% \\
    ILPL  &   Long Parameter List      &  2,117     &   9.84\%     &  13  &  0.31\% \\
    ILI   &   Long Identifier          &  6,841     &   31.81\%    &  12  &  0.29\% \\
    ILS   &   Long Statement           &  11,947    &   55.55\%    &  40  &  0.96\% \\
    IMN   &   Magic Number             &  25,264    &   117.46\%   &  36  &  0.86\% \\
    IMD   &   Missing Default          &  931       &   4.33\%     &  17  &  0.65\% \\

    IVMCC &   Virtual M. C. C.**  &   35    &   0.16\%             &  5   &  0.12\% \\ \hline

    Std Dev &  \multicolumn{1}{c|}{\graycell}  & 7,425.91  &  \multicolumn{1}{c|}{\graycell}  & 11.87 & \multicolumn{1}{c}{\graycell} \\ \hline
    
    Average &  \multicolumn{1}{c|}{\graycell}  & 4,579.36   & \multicolumn{1}{c|}{\graycell}             & 14.63   & \multicolumn{1}{c}{\graycell} \\ \hline

    \multicolumn{6}{l}{**Virtual Method Call from Constructor} \\
    \hline
    \end{tabular}
  \label{tab:implementation_smells}
\end{table}

ILS occurs  when there is an excessively long statement. Long declarations tend to make it difficult to manage the code and are consequently villains if observed from the practice of software testing. Very long code snippets tend to be harder to test because they often become too complex when compared to smaller snippets that are managed more efficiently. 

Finally, IMN occurs when an unexplained number is used in an expression. In general, magic numbers are unique values that have some symbolic meaning. Good programming practices indicate that in these cases, such numbers should be declared as constants to facilitate the reading of the source code, as well as to standardize its use.

Non-VR projects again presented a lower occurrence rate. The most frequent smells were ILS, IMN and IMD which achieved, respectively, percentages of 0.96\%, 0.85\%, and 0.65\%. 

The occurrence of this type of smells is connected with the lack of guidelines for standardization of code as well as the lack of code refactoring practices.  Usually, numbers have a meaning, therefore it is recommended that it should be assigned variables to make the code more readable and self-explaining. The names of the variables should at least reflect what the variable means, not necessarily its value.

Basic guides guide to the test to give a more appropriate context and explanation of whatever numbers are present within the test. The more sloppily the tests are written, the worse the actual code will be and could become a door to possible faults, details about that possibility will be addressed in the next section.

From the standpoint of software testing, opting to use of constants instead of magic numbers can ensure that once the value of the constant has been tested, there is no risk that the value of the constant is erroneously declared in the future.

We also applied the Mann-Whitney test to verify whether there is a statistical difference between the presence of implementation smells in groups of classes that were not tested when compared to the classes that were tested. Adopting a confidence interval of 95\%, the test presented the p-value = 0.00040, which rejects the null hypothesis of the test and confirms the data presented in Table \ref{tab:implementation_smells}, proving that classes that were tested tend to present a lower rate of implementation smells.

\begin{table}[ht]
  \scalefont{0.65}
  \centering
  \caption{Description of the detected design smells and their distribution}
    \begin{tabular}{l|l|r|r|r|r}
    \hline \textbf{ID} & \textbf{Smell} & \multicolumn{2}{c|}{\textbf{VR}} & \multicolumn{2}{c}{\textbf{Non-VR}} \\ \hline
    
    DBH   &   Broken Hierarchy              & 245    & 1.14\%   & 8    & 0.19\% \\
    DBM   &   Broken Modularization         & 991    & 4.61\%   & 46   & 1.10\% \\
    DCM   &   Cyclically-dependent M.       & 3,149  & 14.64\%  & 45   & 1.08\% \\
    DCH   &   Cyclic Hierarchy              & 6      & 0.03\%   & 4    & 0.10\% \\
    DDH   &   Deep Hierarchy                & 0      & 0.00\%   & 1    & 0.02\% \\
    DDE   &   Deficient Encapsulation       & 8,101  & 37.67\%  & 13   & 0.31\% \\
    DDA   &   Duplicate Abstraction         & 2,469  & 11.48\%  & 11   & 0.26\% \\
    DHM   &   Hub-like Modularization       & 4      & 0.02\%   & 16   & 0.38\% \\
    DIA   &   Imperative Abstraction        & 627    & 2.92\%   & 9    & 0.22\% \\
    DIM   &   Insufficient Modularization   & 1,171  & 5.44\%   & 44   & 1.05\% \\
    DMH   &   Missing Hierarchy             & 18     & 0.08\%   & 2    & 0.05\% \\
    DMA   &   Multifaceted Abstraction      & 209    & 0.97\%   & 2    & 0.05\% \\
    DMH   &   Multipath Hierarchy           & 1      & 0.00\%   & 2    & 0.05\% \\
    DRH   &   Rebellious Hierarchy          & 389    & 1.81\%   & 9    & 0.22\% \\
    DUE   &   Unexploited Encapsulation     & 15     & 0.07\%   & 2    & 0.05\% \\
    DUH   &   Unfactored Hierarchy          & 483    & 2.25\%   & 7    & 0.17\% \\
    DUA   &   Unnecessary Abstraction       & 3,741  & 17.39\%  & 17   & 0.41\% \\
    DTA   &   Unutilized Abstraction        & 6,987  & 32.49\%  & 23   & 0.55\% \\
    DWH   &   Wide Hierarchy                & 64     & 0.30\%   & 5    & 0.12\% \\ \hline
    
    Std Dev &  \multicolumn{1}{c|}{\graycell}  & 2,344.41  &  \multicolumn{1}{c|}{\graycell}  & 14.59 & \multicolumn{1}{c}{\graycell} \\ \hline
    
    Average &  \multicolumn{1}{c|}{\graycell}  & 1,508.94   & \multicolumn{1}{c|}{\graycell}             & 14.00   & \multicolumn{1}{c}{\graycell} \\ \hline
    
    \end{tabular}
  \label{tab:design_smells}
\end{table}

\subsubsection{Design smells results}

Finally, we have the design smells which seek to identify breaches of design principles. It can be concluded from Table \ref{tab:design_smells} it is possible to conclude that this class of smells was the one that presented the highest degree of incidence in the VR projects. DUA, DTA and DDE smells were the ones with the highest percentage of occurrence with 17.39\%, 32.49\%, and 37.67\% respectively.

The DUA smell deals with the practice of unnecessary abstractions and is identified when an abstraction has more than one responsibility attributed to it. This smell tends to occur when there is an application of procedural programming features in the context of object-oriented programming languages \cite{Suryanarayana:2014}. 

From the standpoint of VR applications that adopt component-based programming, the appearance of this smell can be explained by the fact that the programming approach focuses on creating interchangeable code modules that work almost independently, not requiring that to be familiar with their inner workings in order to use them.

Unnecessary design abstractions increase their complexity needless and affect the comprehensibility of the overall design. From a software testing point of view, this bad practice tends to hamper test practices

DTA occurs when an abstraction is left unused, is not being used directly, or because it is not reachable in the source code. This smell correlates with DUA since unnecessary abstractions tend not to be used. Another impact factor for the appearance of this smell is linked to possible code maintenance/refactoring activities, which tend to leave traces of code that are no longer needed.

From the standpoint of software testing, the existence of a test base that can be used as a regression test tends to facilitate the localization of source code that is no longer necessary, causing the occurrence of this smell to be reduced. From a tester's point of view, if there is a code that is not being used in the project, it does not need to be tested. Therefore, identifying these snippets of code can lead to more efficient testing activities.

Finally, smell DDE, which identifies cases of poor encapsulation, had the highest occurrence rate in this class of smells. This smell occurs when the declaration of attribute visibility of a class is more permissive than necessary. For example, when the attributes of a class are unnecessary declared as public.

From the standpoint of software testing, separation of interests allows implementation details to be hidden. If an abstraction exposes implementation details unnecessarily, it leads to undesirable coupling in the source code. This will have an impact on the testing activity because checking units that have a high degree of coupling becomes a more challenging task due to the need for more complex mocks and stubs.. Similarly, the high degree of coupling causes changes that are made in a code snippet to reflect in various parts of the application causing previously designed tests to fail if they are not adequately designed.

Non-VR applications had a lower occurrence in this category of smells, in which DBM and DCM are the two that presented the highest occurrence, with 1.10\% and 1.08\% respectively. The explanations for the occurrence rate for the DCM smell are related to the cyclic dependence issue of the MVC model and the DBM smell arises when data and/or methods that ideally should have been localized into a single abstraction are separated and spread across multiple abstractions.

Once again, we applied the Mann-Whitney test to check whether the data obtained from our empirical evaluation can draw a real picture about the behavior  of class that does not have tests compared to classes that were properly tested. Once again the Mann-Whitney test proved with a p-value = 0.00089 that classes that were tested tend to present lower rates of smells when compared to classes that were not tested.

Our second research question ($RQ_2$) sought to understand \textit{``What are the distribution of architecture, implementation and design smells in VR projects?"}. We investigated the main types of smells for VR applications and compared their results with Non-VR applications. We observed that in the context of VR applications there is a greater incidence of code smells related to implementation and design respectively since these two categories have code smells that are repeated frequently due to the characteristics existing in the development of VR applications. 

We also presented a discussion about how software testing practices can benefit from avoiding the smells that obtained the highest occurrence rate. Finally, the statistical tests performed showed that when comparing VR and Non-VR projects, it was possible to observe that the software testing practice can contribute to increase the quality criteria and to reduce the presence of code smells.

\begin{figure*}
    \centering
    \includegraphics[scale= 0.8]{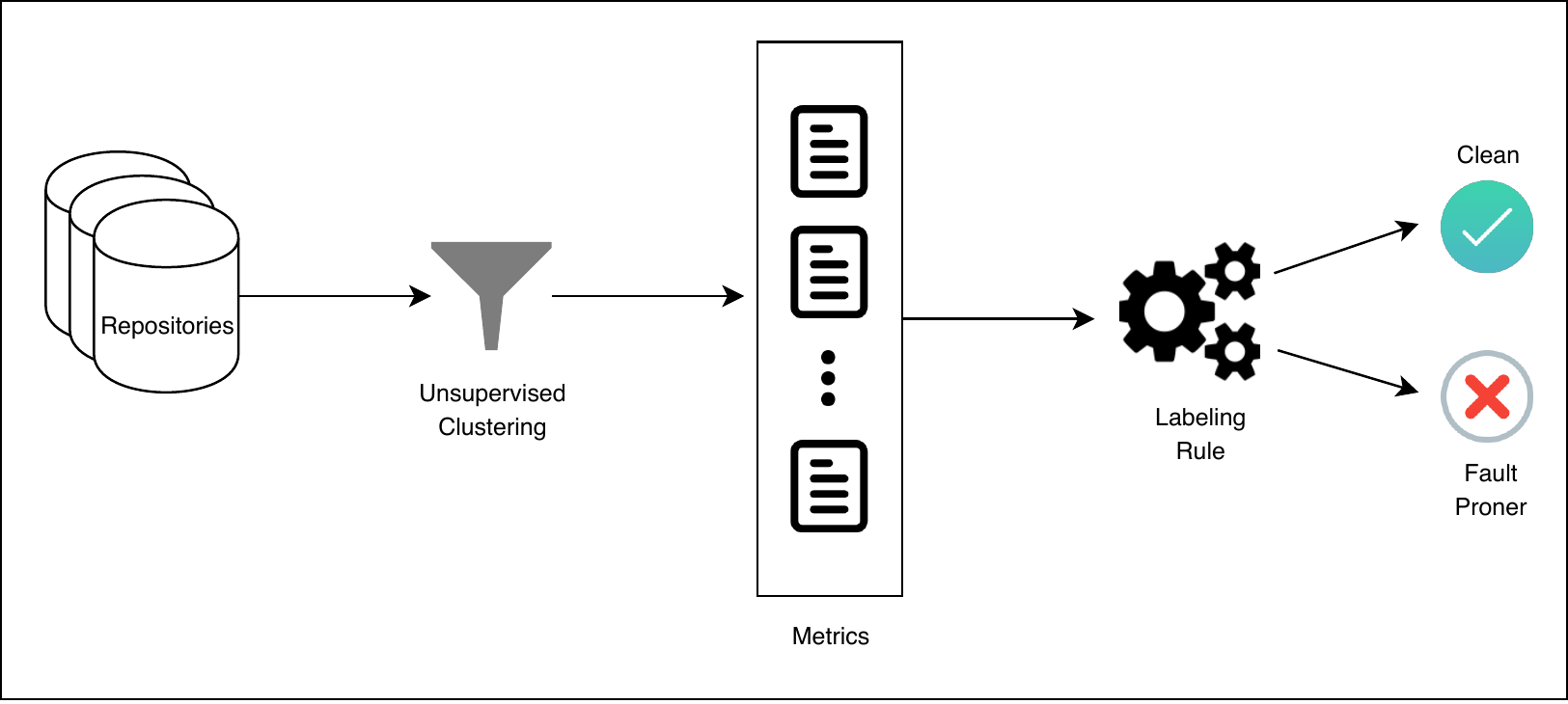}
    \caption{General process of ACL fault prediction approach (Source: own construction)}
    \label{fig:acl_workflow}
\end{figure*}

According to Hall et al. \cite{Hall:2014}, code smells have a significant but small effect on faults. This can justify the fact that Non-VR application projects, which have test cases, present a lower rate of code smells when compared to VR applications, which do not have, for the most part, a well-defined test activity.  However, the presence of smells not only hides potential source code flaws but also contributes to hindering the maintainability and evolution of the source code in larger projects. This leads us to the last research question of this study ($RQ_3$), which aims to investigate the fault proneness of VR  projects. In a similar way to the analysis of code smells, evaluate this in more depth, we inserted the analysis of Non-VR projects so as to better understand the results.

\subsection{Analyzing fault proneness}

As mentioned before, the presence of code smells can indicate the absence of quality attributes in the source code and this can be an indication of faults in a software \cite{Hall:2014}. Similarly, as previously mentioned, the higher occurrence rate of code smells in the projects can hinder the practice of software testing. To understand the risks of neglecting this activity, we analyzed the projects concerning fault proneness.

Since code smells are identified according to rules and thresholds defined in code metrics \cite{Khomh:2009}, we aim to investigate ($\mathbf{RQ_3}$) the question: \textit{``Can we draw a relationship between code metrics and fault proneness?"}. To do so, we use the code metrics described in Table \ref{tab:projects_metrics} with a fault prediction technique, which uses the metrics value as an indicator to suggest whether a given source code is fault-prone or not.

By exploring relationships between software metrics and fault proneness, we seek to justify the need for software testing activities. For instance, a high threshold in a specific metric may lead us to suspect, with high probability, about the reliability of some parts of the code. 

The effectiveness of fault prediction techniques is often demonstrated using historical repository data \cite{Herbold:2017}. However, once these techniques are adopted, it is not clear how they would affect projects that do not match with the characteristics (language, platform, domain) of the built model \cite{Peng:2015}.

Since we do not have access to a dataset or a bug track history maintained with VR systems data, we tried to exploit an approach that uses an unsupervised fault prediction technique \cite{Yang:2016}, that does not rely on historical data, to investigate fault proneness on the analyzed projects.    

We use the Average Clustering and Labeling (ACL) \cite{Yang:2016} approach to predict fault proneness in unlabeled datasets. ACL models obtain good prediction performance and are comparable to typical supervised learning models in terms of precision and recall, offering a viable choice for fault prediction when we do not have historical data related to faults.

This study can help software developers to understand the characteristics of VR software and the potential implications of neglect software testing activities. Raising awareness is the first step towards VV\&T activities.

Figure \ref{fig:acl_workflow} describes the process used by the approach to attest if a given instance of code is defined as fault-prone or not. In general terms, the the process to run the approach consist mostly of four steps:

\begin{itemize}
    
    \item calculates the average value for each of the code metrics used;
    
    \item build a violation matrix metric;
    
    \item calculates metrics of instance violation; and
    
    \item defines whether the analyzed instance is considered as fault-prone or clean.
    
\end{itemize}

The first step is self-explanatory and uses the individual metrics for each class, similar to those presented in Table \ref{tab:projects_metrics}, as input data. After this stage, a violation matrix, which evaluates each metric, is built using as a basis the average of the results constituted for all classes in the project. The next step is to check the number of violations for each class concerning the metrics evaluated and, finally, in the last step to classify whether a particular instance is fault-prone or not.

To perform the classification, it is necessary to define a cutoff that will be used as a threshold and, if violated, it will identify whether the class is fault-prone or not. The cutoff point is calculated using the number of metrics that are used in the evaluation. Full details of the approach and the implementation can be found in \cite{Yang:2016} and in the repository that contains the information about this work.

The 119 VR projects were analyzed using the described approach and according to the classification metric adopted, from 21,508 classes contained in all the projects, a total of 2,627 classes or 12.21\% were classified as classes with a high probability of having faults, due to the fact they extrapolate the threshold defined by the approach to consider them as clean.

Similarly, in the 107 Non-VR projects, out of 21,568 classes, a total of 1,921 were labeled as fault-prone, which corresponds to a percentage of 8.90\% of the analyzed classes.

In a superficial analysis it is possible to have a mistaken view of the results of this study and imagine that the percentage of propensity to fail presented appears low and, therefore, the time and expense necessary to identify them is perhaps not justified. However, according to previous investigations \cite{Walkinshaw:2018}, the \textit{Pareto} principle also tends to apply to a software faults context. It is believed that 20\% of the files are responsible for up to 80\% of the faults found in a project. Therefore, it is natural that the percentage of classes with a propensity to fail to follow this same trend, since it is not an exact proportion and the classification approach is not an exact formula and serves only as a mechanism to assist in efforts to apply a software testing approach.

As pointed out by Nam \cite{Nam:2014}, defect prediction approaches play a role as a complementary approach to help identify potential problems in the source code as well as a mechanism to improve it and consequently get rid of productivity bottlenecks and future issues. Thus, the results presented here are not intended to point out the exact number of problems in a software product evaluated, but to strengthen the hypotheses that projects that adopt quality criteria, such as software testing practice, tend to be less predisposed to future issues.

It's also important to note that, since there is no precise information about the test criteria used in Non-VR projects, as well as any information regarding the coverage reached by the designed tests, it is impossible to guarantee that the tests designed for a class are enough to ensure  that it is free of any problems. Therefore, it is natural that the percentage of fault-proneness between projects that have not been tested (VR projects) and projects that have test cases (Non-VR projects) is slightly similar.

It can be observed that despite having a larger number of classes and lines of code for those analyzed in the VR projects, Non-VR projects presented a lower fault-proneness rate. It is worth noting that the fault-prone algorithm is executed only in the classes related to the source code of the application, thus disregarding the test classes in the Non-VR projects. 

This analysis could be an indication that due to the practice of testing, classes of the Non-VR projects have a higher degree of reliability, and therefore are less fault-prone when compared to the classes existing in the VR projects, which mostly do not present test cases.

\begin{figure*}
    \centering
    \includegraphics[scale = 0.35]{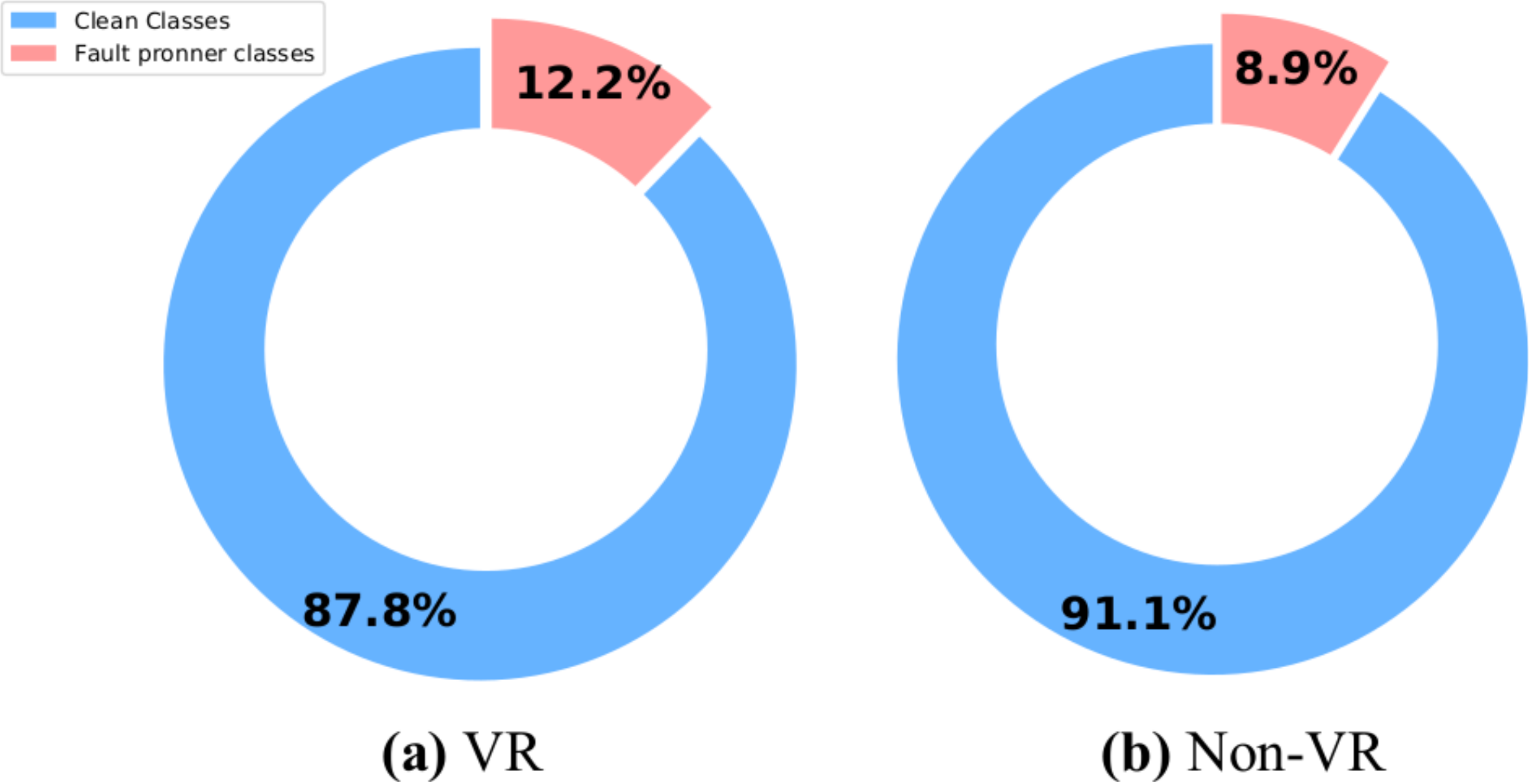}
    \caption{Classification of the VR projects according to the ACL approach (Source: own construction)}
    \label{fig:pie_chart_vr}
\end{figure*}

These numbers can be observed in Figure \ref{fig:pie_chart_vr} and are alarming numbers since they show the negative impact that a lack of robust and standardized testing technologies can cause to the software industry \cite{Planning:2002}. Consequently, it contributes to an increase in the incidence of avoidable faults that tend to appear only after the software is used by its end users.

Similarly, the software development cost tends to increase because historically the process of identifying and correcting faults during the software development process represents more than half of the costs incurred during the development cycle \cite{Brooks:1995}. This delay in the product development can lead to situations such as the increase in the time needed to put a product on the market, also resulting in market opportunity losses \cite{Afonso:2008}.

We went further trying to understand how the faults pointed out by the approach are distributed into the projects. Since the projects have a great variety of sizes, we grouped them into 6 different categories (by the number of classes) in order to observe how the distribution of fault-prone classes occurs.

\begin{figure}[ht]
    \centering
    \includegraphics[scale=0.6]{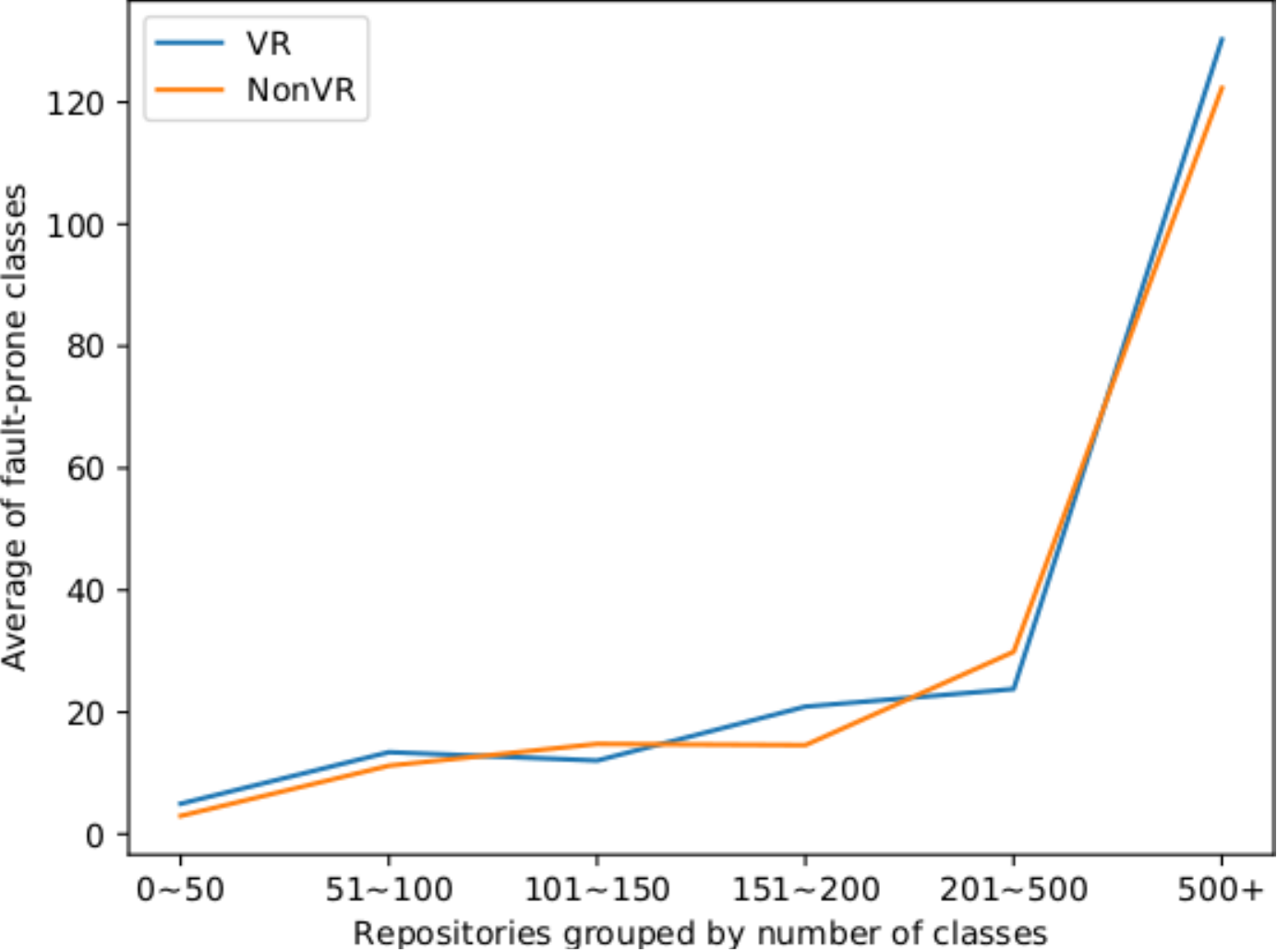}
    \caption{Distribution of fault-prone classes according to the size of the projects (Source: own construction)}
    \label{fig:lines_chart}
\end{figure}

Figure \ref{fig:lines_chart} shows this distribution. It can be observed that in both VR and Non-VR projects, there is a relation between the number of fault-prone classes and the size of the projects. This relation points out that the higher the number of classes in the projects, the higher the average fault-prone classes, and leads us to conclude that neglecting testing activity in larger projects may be  even more riskier in terms of the project’s success.

Future analyses could be extracted from the data obtained. However, we believe that the presented data are capable of attesting a clear answer to $RQ_3$, making it clear that in a general context, the lack of software testing techniques have a direct impact on quality attributes, as demonstrated by the metrics extracted from the analyzed projects and this directly reflects the adoption of bad development practices, which lead to the existence of code smells, consequently becoming an outlet for the increase in faults.

$RQ_3$ sought to understand the question: \textit{``Can we draw a relationship between code metrics and fault proneness?"} and by implementing the approach to detect fault proneness in the projects investigated, it could be observed that neglecting the test activity can lead to a higher probability of development problems. According to our analysis, it was seen that the VR  projects, which do not present test cases, have a higher propensity to present faults to Non-VR projects, and that propensity tends to increase as the complexity of the projects increases.

It was also observed that although Non-VR projects present test cases in all projects, they still present a high rate of fault proneness. This underscores the importance of the developing software testing practice within the scope of project development. Although Non-VR projects have test cases, the test sets provided do not meet the basic test criteria, such as code coverage, so that part of the code that is not tested is still prone to possible failures.

Another point that this study raised is the need for specific test practices for a specific domain. Software of different domains have different characteristics, which must be adequately investigated. In the context of VR applications, the simple use of unit tests may not be sufficient to attest the quality of the developed product, since the technological advancement has led to the development of systems with advanced features such as images, sounds, videos, and differentiated interaction, presenting new challenges when compared to software testing in conventional domains, such as the lack of information on typical defects and even the lack of a precise definition of a test case and the oracle problem \cite{Rapps:1985, Barr:2015}.

\section{Limitations and Threats to Validity}\label{sec:limitations}

The main limitations of this study are related to the fact that the data used in this study were gathered from \textit{Github}. 

Although the collected data enabled us to discuss the state of practice regarding the application of software testing techniques in the context of VR, open source projects still represent only a small portion of what is produced in the context of VR applications. Commercial projects and closed projects are also part of this universe, and it is not possible to attest that the results discussed from data extracted from open source projects can be generalized for these other scenarios.

The problem described above opens opportunities to develop similar research in partnership with industry in order to understand whether the results converge in the same direction.

In addition to the limitations described above, another obstacle that must be pointed out is the fact that despite the fact that the assumptions made during the study were related to the context of VR applications, it should be emphasized that all the samples observed only use a single technology (\textit{Unity}), therefore the results indicated here cannot be generalized for other platforms. To achieve such generalization, new studies should be carried out to corroborate or counter the results presented in this study.

The validity of the results achieved in experiments depends on factors in the experiment settings. Different types of validity can be prioritized depending on the goal of the experiment. Concerning the threats to the validity of this study, which are related both to the evaluation of code smells and to fault-proneness detection, we can highlight the fact of performing an analysis using samples extracted only from a platform and just for open source projects was a significant threat to validity   - this relies on the lack of representativity of the projects in serving as a real sample of the universe of all types of projects for the VR domain. 

Unfortunately the lack of representativity is a problem that affects the entire software engineering area, since there is no well-fledged theory capable of ensuring that a set of programs is considered a representative sample for experimentation. To try to mitigate this threat, the most significant possible number of projects was assembled, varying in size (small, medium and large) and application purposes (entertainment, simulation, training, health).

Another measure taken to try to mitigate the threat described above was to analyze Non-VR projects, which served as subsidies to compare with the results obtained from VR projects, ensuring a better grounded discussion and a minimum baseline for comparison, since, unfortunately, there are still no projects cataloged with VR applications that meet the requirements to be used in this work.

Related to threats to construct validity, possible mistakes can be pointed out both in the analysis of codes smells, as well as in the evaluation of fault proneness. To minimize this threat, the tool \textit{Designite} was used to detect code smells, \textit{Designite} is a commercial tool and has already been successfully used in other experiments \cite{Sharma:2017}. 

In the context of the experiments carried in this study,  in addition to using a commercial tool to support the gathering of data related to the code smell, all the data presented was evaluated using both descriptive analyses, and hypothesis testing in order to guarantee that the conclusions drawn from this work enable us to paint a clear and quantitative picture about the subject explored.

Regarding the approach to detect fault proneness, the strategy used was previously validated through experiments in large datasets to attest its efficacy \cite{Yang:2016}, and it is worth mentioning the fact that the main point of the approach was not, in fact, to detect faults in the projects, but point out classes that have a high probability of having faults, serving as a guide to direct testing efforts and to discuss the necessity of apply software testing techniques.

Finally, we discuss threats to the internal validity of the study, which are related to the level of confidence between the expected results and the results obtained. The whole study was conducted in a way that minimized this threat. To increase confidence about the presented results, the data were analyzed using tables and graphs and were also made available in a repository to enable the replication if it is deemed necessary.

\section{Conclusion and Future Work}\label{sec:conclusion}

This paper discusses the main challenges related to using software testing practice in the VR domain. Some of the critical issues related to the quality of these systems were pointed out and possible solutions were also discussed that could be used and adapted to deal with such issues.

We discussed whether or not there is a real need to test VR systems. To better understand this, a comprehensive study was conducted, guided by 3 research questions, whose objective was: to understand the state of the practice of software testing in the context of VR programs ($RQ_1$), to measure metrics and quality attributes in VR software ($RQ_2$), and finally to evaluate fault proneness in the collection of the software analyzed ($RQ_3$).

In order to answer the raised questions, a collection of 119 VR projects, available in open source projects and manually analyzed, was cataloged to understand the state of the practice concerning the application of software testing techniques. Regarding the application of software testing techniques ($RQ_1$), it was observed that out of all the projects, only 6 of them had some test cases in their project.

Given the results pointed out by $RQ_1$, we decided to evaluate how the negligence of the practice of software testing can be detrimental to a software project, and it was decided to evaluate the distribution of code smells among the analyzed projects. Smells related to architecture, design, and implementation were analyzed. It can be concluded that there is a high incidence of smells in the projects analyzed, especially regarding implementation smells. We discussed the most common smells for each of the categories and how they can discourage the practice of software testing, and also how they can be avoided if a software testing activity is appropriately conducted.

Finally, considering the results of $RQ_2$, it was decided to investigate how the lack of good practices and the presence of code smells can impact the quality of the source code produced. To do so, an approach that evaluates code metrics was used to point out classes that are fault-prone ($RQ_3$). The study pointed out that about 12\% of the analyzed VR classes have such characteristics, revealing a significant risk to the success of the projects. The distribution of these classes was also evaluated when observed concerning the size of the projects analyzed. It was observed that the larger a project becomes, there is a higher incidence of fault-prone classes, which may be an indication that neglecting test practices in larger projects becomes even more riskier.

We believe that the results reported in this paper will contribute to raising the awareness of the software testing and virtual reality community about the needs of software testing approaches for VR developers. As software testing phase, also makes up one of the development phases, it is necessary to understand the point of view of stakeholders involved in the process, allowing these groups narrow what they deem important, thus making it possible to prioritize verification concerning failures that should not manifest in VR applications.

Observing such aspects, it is possible to guide the development of a project using a specific testing approach for the VR domain. So, in future works, we intend to survey what types of faults in VR applications contribute to a negative experience. The goal of this study is to obtain a view of the interest groups (for example, which types of failures are most critical, which are less relevant, how much each affects the quality of the final product, etc.), in addition to investigating the knowledge of the groups of interest regarding specific types of failures in VR applications.

Understanding the intends of stakeholders we expect to propose a fault taxonomy to the context of VR programs. It is believed that having such taxonomy, it would be possible to encourage the development of specific software testing techniques and criteria to the context of VR programs, thus spreading the practice of software testing in order to mitigate possible problems and move towards software projects that best meet software quality requirements.

\section*{Acknowledgements}

Stevão A. Andrade research was funded by FAPESP (São Paulo Research Foundation), process number 2017/19492-1. This study was also financed in part by FAPESP (São Paulo Research Foundation) process number 2019/06937-0. The authors are grateful to Brazilian National Council of Scientific and Technological Development (CNPq) for assistance (process 308615/2018-2), and National Institute of Science and Technology Medicine Assisted by Scientific Computing INCT MACC ( Process 157535/2017-7).

We also would like to thanks Tushar Sharma and \textit{Designite} team by providing us an Academic license of \textit{Designite} tool.

\bibliographystyle{apalike}
\bibliography{main.bib}

\end{document}